\input{aipcheck}

\documentclass[
    final            % use final for the camera ready runs
  ]
  {aipproc}

\layoutstyle{6x9}

\begin{document}

\title{Glueball-glueball scattering in a constituent gluon model}

\author{M\'ario L. L. da Silva}{
  address={Instituto de F\'{\i}sica, Universidade Federal do Rio Grande do
  Sul, CEP 91501-970, Porto Alegre, Rio Grande do Sul, Brazil}
%  email={mariolls@if.ufrgs.br},
%  thanks={This work was commissioned by the CNPq}
}

\author{Dimiter Hadjimichef}{
  address={Instituto de F\'{\i}sica e Matem\'atica, Universidade Federal de
  Pelotas, CEP 96010-900, Pelotas, Rio Grande do Sul, Brazil}
%  email={dimiter@ufpel.edu.br},
}

\author{C\'esar A. Z. Vasconcellos}{
  address={Instituto de F\'{\i}sica, Universidade Federal do Rio Grande do
  Sul, CEP 91501-970, Porto Alegre, Rio Grande do Sul, Brazil}
%  ,altaddress={Instituto de F\'{\i}sica, Universidade Federal do Rio Grande do
%  Sul, CEP 91501-970 Porto Alegre, Rio Grande do Sul, Brazil} 
  % additional visiting address
%  email={mariolls@if.ufrgs.br, cesarzen},
%  thanks={This work was commissioned by the CNPq}
}

\begin{abstract}
In this work we use a mapping technique to derive in the
context of a constituent gluon model an effective
Hamiltonian that involves explicit gluon degrees of freedom.
We study glueballs with two gluons using the Fock-Tani formalism.
In the present work we consider two possibilities for $0^{++}$:
({\it i}) as a pure $s\bar{s}$ and calculate, in the context of 
a quark interchange picture, the cross-section; ({\it ii} ) as 
a glueball where a new calculation for this cross-section is made, 
in the context of the constituent gluon model, with gluon interchange.
\end{abstract}

\maketitle

%%%%%%%%%%%%%%%%%%%%%%%%%%%%%%%%%%%%%%%%%%%%
%% MAINMATTER
%%%%%%%%%%%%%%%%%%%%%%%%%%%%%%%%%%%%%%%%%%%%

\section{Introduction}

The gluon self-coupling in QCD implies the existence of bound states
of pure gauge fields known as glueballs. Numerous technical
difficulties have so far been present in our understanding of their
properties in experiments, largely because glueball states can mix
strongly with nearby $q\bar{q}$ resonances. However recent
experimental and lattice studies of $0^{++}$, $2^{++}$ and $0^{-+}$
glueballs seem to be convergent.

%  On theoretical grounds, a 
%simple potential model with massive constituent gluons, namely the 
%model of Cornwall and Soni [1],[2] has been  studied [3],[4] and 
%the results are consistent with lattice and experiment.

In this work we use a mapping technique to derive in the context 
of a constituent gluon model an effective Hamiltonian that involves 
explicit gluon degrees of freedom. We study the glueball  as a 
bound-state of  two constituent gluons using the Fock-Tani formalism \cite{annals}.
A glueball-glueball potential $V_{gg}$, scattering amplitude $h_{fi}$
and a  cross-section $\sigma_{gg}$ can be obtained.

In the conventional quark model a $0^{++}$ state with $M_{0^{++}}=1.73$ GeV
is considered as $q\bar{q}$ bound state. The $0^{++}$ resonance is a isospin zero 
state so, in principal, it can be either represented as a  
$q\bar{q}$  bound state, a glueball, or a mixture. In particular
there is growing evidence in the direction of large $s\bar{s}$ content
with some mixture with the glue sector. 
It turns out that this resonance is an interesting system, in the
theoretical point of view, where one can compare models. In the
present work we consider two possibilities for $0^{++}$: ({\it i})
as a pure $s\bar{s}$ and calculate, in the context of a quark 
interchange picture, the cross-section; ({\it ii} ) as a glueball
where a new calculation for this cross-section is made, in the 
context of the constituent gluon model, with gluon interchange.

On theoretical grounds, a simple potential model with massive 
constituent gluons, interacting by a two-body potential $V_{\rm aa}$, 
namely the model of Cornwall and Soni
\cite{cs1,cs2} is used as an input in the Fock-Tani formalism for the microscopic
model. This model  has been studied recently in \cite{cs3,cs4} 
and the results are consistent with lattice and experiment. 
A glueball-glueball potential can be obtained applying in a 
standard way the Fock-Tani transformed operators to the 
microscopic Hamiltonian
  \begin{eqnarray*}
    {\cal{H}}(\mu\nu;\sigma\rho)=T_{\rm aa}(\mu)a_{\mu}^{\dagger}a_{\mu}+\frac{1}{2}
    V_{\rm aa}(\mu\nu;\sigma\rho)a_{\mu}^{\dagger}a_{\nu}^{\dagger}a_{\rho}a_{\sigma}
  \end{eqnarray*}
  where one obtains for the glueball-gluball potential $V_{gg}$
  \begin{eqnarray}
    V_{gg}=\sum_{i=1}^{4}V_{i}(\alpha\gamma;\delta\beta)g_{\alpha}^{\dagger}
    g_{\gamma}^{\dagger}g_{\delta}g_{\beta}
\label{v_gg}
  \end{eqnarray}
  and the $V_{i}(\alpha\gamma;\delta\beta)$ are given by
\begin{eqnarray*}
   & &V_{1}=2V_{aa}(\mu\nu;\sigma\rho)\Phi_{\alpha}^{\star\mu\tau}
    \Phi_{\gamma}^{\star\nu\xi}\Phi_{\delta}^{\rho\xi}\Phi_{\beta}^{\sigma\tau}
\,\,\,\,,\,\,\,\,
    V_{2}=2V_{aa}(\mu\nu;\sigma\rho)\Phi_{\alpha}^{\star\mu\tau}
    \Phi_{\gamma}^{\star\nu\xi}\Phi_{\delta}^{\rho\tau}\Phi_{\beta}^{\sigma\xi}\\
   &&V_{3}=V_{aa}(\mu\nu;\sigma\rho)\Phi_{\alpha}^{\star\mu\nu}
    \Phi_{\gamma}^{\star\lambda\xi}\Phi_{\delta}^{\sigma\lambda}\Phi_{\beta}^{\rho\xi}
\,\,\,\,,\,\,\,\,
    V_{4}=V_{aa}(\mu\nu;\sigma\rho)\Phi_{\alpha}^{\star\mu\xi}
    \Phi_{\gamma}^{\star\nu\lambda}\Phi_{\delta}^{\lambda\xi}\Phi_{\beta}^{\rho\sigma}\,.
\end{eqnarray*}

\begin{figure}[ht]
%\begin{center}
\scalebox{.45}{\includegraphics*[10pt,600pt][580pt,745pt]{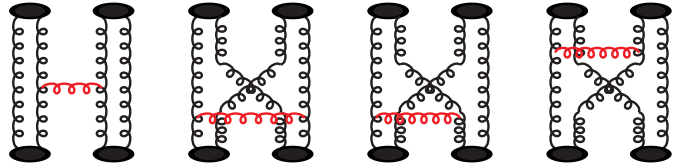}}
%\end{center}
\caption{Diagrams representing the scattering amplitude $h_{fi}$
for glueball-glueball interaction with constituent gluon
interchange.}
\end{figure}

The next step is to obtain the  scattering $T$-matrix from Eq.
(\ref{v_gg}). Due to translational invariance, the $T$-matrix element is
written as a momentum conservation delta-function, times a
Born-order matrix element, $h_{fi}$:  
\begin{eqnarray*}
T(\alpha\beta;\gamma\delta)=(\alpha\beta|V_{gg}|\gamma\delta)=\delta^{(3)}(
\vec{P}_{f}-\vec{P}_{i})\,h_{fi} 
\end{eqnarray*} 
 where $\vec{P}_{f}$ and $\vec{P}_{i}$ are the final and initial momenta of the
two-glueball system.  This result can be used in order to evaluate
the glueball-glueball scattering cross-section 
\begin{eqnarray} 
\sigma_{gg}
=\frac{4\pi ^{5}\,s}{s-4M_{G}^{2}}
\int_{-(s-4M_{G}^{2})}^{0}\,dt\,|h_{fi}|^{2} 
\label{cross} 
\end{eqnarray}
where $M_G$ is the glueball mass, $s$ and $t$ are the Mandelstam
variables.

\begin{figure}[ht]
%\begin{center}
\scalebox{.45}{\includegraphics*[50pt,50pt][520pt,450pt]{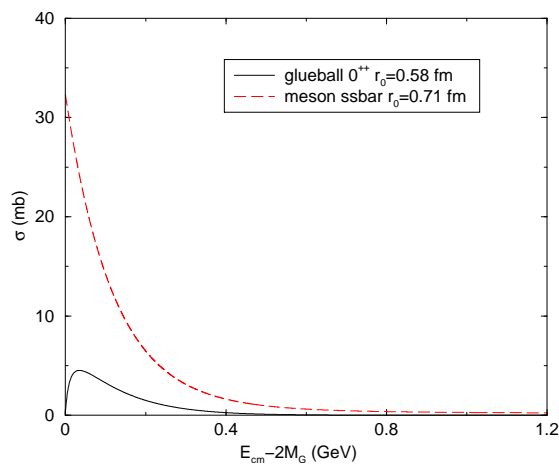}}
%\end{center}
\caption{Cross-section comparison for $0^{++}$ with the following
parameters $\beta=0.1$, $\lambda = 1.8$, $k=0.21$, gluon mass
$m=0.6$ GeV. The  $s\bar{s}$ quark model parameters: $m_{q}=0.55$ GeV,
$\alpha_{s}=0.6$.} 
\end{figure}
%\vspace{-0.5cm}
\section{Conclusions}

In this work we have extended the Fock-Tani formalism to a
hadronic model in which the bound state is composed by bosons (gluons). The
Cornwall-Soni constituent gluon model was used as the microscopic
model.  This model depends on
several parameters: $\lambda=3\alpha_s$; $\beta$, the string tension; $k$, a smearing constant and
$m$, the constituent gluon mass. In the present we solve a mass equation in order to fix these parameters:
 $\beta=0.1$, $\lambda = 1.8$, $k=0.21$, gluon mass $m=0.6$ GeV, for a radius $r_0=0.58$ fm.
 A comparison of the cross-sections reveals that a $0^{++}$ composed by  $s\bar{s}$ quarks, with parameters 
$\alpha_s=0.6$, $m_q=0.55$ GeV and $r_0=0.71$ fm,  adjusted to fit $\pi-\pi$ 
and $K-K$ scattering \cite{annals} implies in a
much larger cross-section and different in  shape than in the constituent gluon picture.
This could represent a guiding-line criterion for distinguishing between
pictures. The $0^{++}$ in  our calculation is an extreme case, in which one has two models
representing a pure state resonance. For other resonances, with a larger degree of mixture, 
this effect in the  cross-section will not be clean, but we believe it will still be present validating
this approach as a form of identifying glueballs in  the meson spectrum.
\vspace{-0.3cm}
\begin{theacknowledgments}
The author (M.L.L.S.)  was supported by Conselho Nacional de
Desenvolvimento Cient\'{\i}fico e Tecnol\'ogico (CNPq).
\end{theacknowledgments}

\end{document}